\documentclass[aps,prl,superscriptaddress,showpacs,twocolumn]{revtex4}
\usepackage{graphicx}

\begin{document}


\title{Shape Isomerism at $\mathbf{N=40}$: Discovery of a Proton Intruder in $\mathbf{^{67}}$Co}


 \author{D.~Pauwels}
 \email[]{Dieter.Pauwels@fys.kuleuven.be}
 \affiliation{Instituut voor Kern- en Stralingsfysica, K.U. Leuven,
 Celestijnenlaan 200D, B-3001 Leuven, Belgium}
 \author{O.~Ivanov}
 \affiliation{Instituut voor Kern- en Stralingsfysica, K.U. Leuven,
 Celestijnenlaan 200D, B-3001 Leuven, Belgium}
 \author{J.~B\"{u}scher}
 \affiliation{Instituut voor Kern- en Stralingsfysica, K.U. Leuven,
 Celestijnenlaan 200D, B-3001 Leuven, Belgium}
 \author{T.E.~Cocolios}
 \affiliation{Instituut voor Kern- en Stralingsfysica, K.U. Leuven,
 Celestijnenlaan 200D, B-3001 Leuven, Belgium}
 \author{J.~Gentens}
 \affiliation{Instituut voor Kern- en Stralingsfysica, K.U. Leuven,
 Celestijnenlaan 200D, B-3001 Leuven, Belgium}
 \author{M.~Huyse}
 \affiliation{Instituut voor Kern- en Stralingsfysica, K.U. Leuven,
 Celestijnenlaan 200D, B-3001 Leuven, Belgium}
 \author{A.~Korgul}
 \affiliation{Institute of Experimental Physics, Warsaw University, ul.Ho\.{z}a 69, 00-681 Warszawa, Poland}
 \author{Yu.~Kudryavtsev}
 \affiliation{Instituut voor Kern- en Stralingsfysica, K.U. Leuven,
 Celestijnenlaan 200D, B-3001 Leuven, Belgium}
 \author{R.~Raabe}
 \affiliation{Instituut voor Kern- en Stralingsfysica, K.U. Leuven,
 Celestijnenlaan 200D, B-3001 Leuven, Belgium}
 \author{M.~Sawicka}
 \affiliation{Instituut voor Kern- en Stralingsfysica, K.U. Leuven,
 Celestijnenlaan 200D, B-3001 Leuven, Belgium}
 \author{I.~Stefanescu}
 \affiliation{Instituut voor Kern- en Stralingsfysica, K.U. Leuven,
 Celestijnenlaan 200D, B-3001 Leuven, Belgium}
 \affiliation{Department of Chemistry and Biochemistry, University of Maryland, College Park, Maryland 20742, USA}
 \author{J.~Van de Walle}
 \affiliation{Instituut voor Kern- en Stralingsfysica, K.U. Leuven,
 Celestijnenlaan 200D, B-3001 Leuven, Belgium}
 \author{P.~Van den Bergh}
 \affiliation{Instituut voor Kern- en Stralingsfysica, K.U. Leuven,
 Celestijnenlaan 200D, B-3001 Leuven, Belgium}
 \author{P.~Van Duppen}
 \affiliation{Instituut voor Kern- en Stralingsfysica, K.U. Leuven,
 Celestijnenlaan 200D, B-3001 Leuven, Belgium}
 \author{W.B.~Walters}
 \affiliation{Department of Chemistry and Biochemistry, University of Maryland, College Park, Maryland 20742, USA}



\date{\today}

\begin{abstract}
The nuclear structure of $^{67}$Co has been investigated through $^{67}$Fe $\beta$-decay. The $^{67}$Fe isotopes were produced at the LISOL facility in proton-induced fission of $^{238}$U and selected using resonant laser ionization combined with mass separation. The application of a new correlation technique unambiguously revealed a $496 (33)$~ms isomeric state in $^{67}$Co at an unexpected low energy of $492$~keV. A $^{67}$Co level scheme has been deduced. Proposed spin and parities suggest a spherical $(7/2^-)$ $^{67}$Co ground state and a deformed first excited $(1/2^-)$ state at $492$~keV, interpreted as a proton $\mathrm{1p-2h}$ prolate intruder state.
\end{abstract}


\maketitle

Atomic nuclei in the neighborhood of closed shells often exhibit intriguingly low-energy excitations whereby particle-hole configurations across major shell gaps give rise to so-called intruder states \cite{Hey_PR_83,Woo_PR_92}. Although expected at an excitation energy of at least the size of the shell gap, the strong energy gain in both pairing and proton-neutron interactions can bring them close to the ground state. In odd-mass nuclei with $\pm$ one nucleon outside a closed shell, the possible unique spin/parity of the intruder orbitals combined with the difference in deformation compared to the normal states can lead to isomerism. Their excitation energy becomes minimal where the proton-neutron correlations are maximal, typically in the middle of the open shell. Intruder states, through their isomeric character, are excellent experimental and theoretical probes to study the relation between individual and collective excitations in atomic nuclei revealing information on shell gaps, pairing correlations and proton-neutron interactions. The rich variety in orbitals, shell gaps and shapes available in exotic nuclei makes intruder states an ideal laboratory for a more general study of mesoscopic systems.

Through the use of a novel correlation technique applied to the $\beta$-decay of $^{67}$Fe, we report in this Letter on the surprising existence of an intruder isomer in $^{67}_{27}$Co$_{40}$ at an excitation energy of $492$~keV. This discovery marks the first time that a $1$-particle-$2$-hole intruder state has been identified so low in energy adjacent to $^{68}$Ni, a nucleus that exhibits ''double-magic'' properties in the form of a spike in the positions of the first $2^{+}$ and $4^{+}$ levels \cite{Bro_PRL_95,Ste_PRL_08,Van_PRC_04,Mue_PRL_99,Wei_PRC_99,Sor_PRL_02}.

In previous $^{67}$Fe decay studies the half-life of $470(50)$~ms was reported \cite{Ame_EPJ_98} and a single gamma ray at $189$~keV was identified \cite{Sor_NPA_00}. Low production yields far away from the line of stability and difficulties to produce the short-lived cobalt and iron isotopes using conventional ISOL techniques hamper detailed studies. However, much improved data for the decay of neutron-rich iron and cobalt nuclei can now be obtained with high selectivity using the laser-ion source at the LISOL facility \cite{Fac_NIM_04}. The $^{67}$Fe nuclei were produced in a $30$~MeV proton-induced fission reaction on two $10 \ \mathrm{mg}/\mathrm{cm}^2$ thick $^{238}$U targets, placed inside a gas cell. The fission products, recoiling out of the targets, were stopped and thermalized in argon buffer gas at a pressure of $500$~mbar. The iron isotopes were selectively laser-ionized by two excimer-pumped dye lasers close to the exit hole of the gas cell. After mass separation the ions were implanted into a tape, which was surrounded by three thin plastic $\beta$-detectors and two MINIBALL $\gamma$-detector clusters \cite{Ebe_NPA_01}. This detection setup combined with the laser-ion source offers the possibility to apply a new correlation method \cite{Pau_NIM_08}.

\begin{figure}[ttt]
\centering
\includegraphics[width=\linewidth]{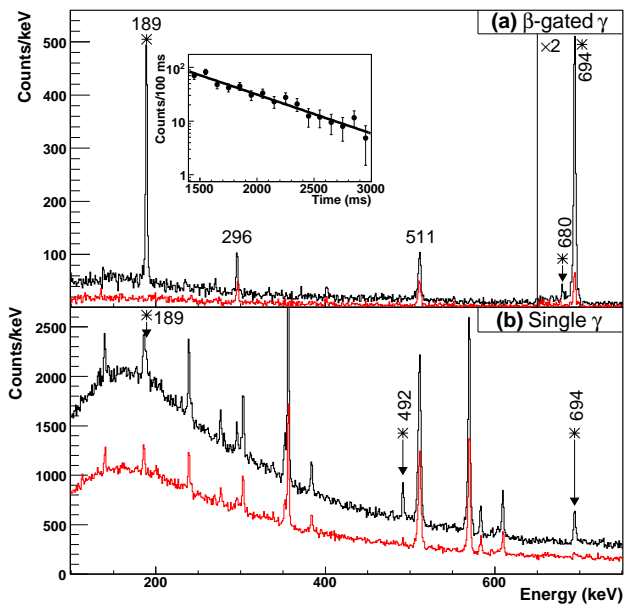}
 \caption{$\gamma$-ray spectra at $A=67$. (a) $\beta$-gated spectra with the laser-on (upper-black on-line) and laser-off (lower-red on-line). (b) Singles $\gamma$-ray spectra with the laser-on (upper-black on-line) and the laser-off (lower-red on-line). Lines marked by a star are laser-enhanced lines and the $296$-keV line is from $^{102}$Nb $\beta$-decay reaching the detection set-up in the doubly charged molecular form $^{102}\mathrm{Nb}\mathrm{O}_{2}^{++}$. The lines that are not marked are background lines. The upper insert in (a) is the fitted decay behavior of the $189$-keV line.}
 \label{fig:Fe_P04}
\end{figure}

In Fig.~\ref{fig:Fe_P04} $\gamma$-spectra are shown, detected in prompt coincidence ($350$-ns window) with $\beta$-particles (''$\beta$-gated'', (a)) or without any coincidence (''singles'', (b)). The black spectra were acquired over $11$~h with the lasers tuned to the resonance frequency for iron, while the red spectra were acquired over $6$~h with the lasers off. By comparing both $\beta$-gated spectra, a number of $^{67}$Fe decay lines could be identified, amongst them the $189$-keV line, observed already in Ref.~\cite{Sor_NPA_00} and the $680$-keV line. The ground-state half-life of $^{67}$Fe was re-measured from the time dependence of the $\beta$-gated $189$-keV $\gamma$-ray as shown by the inset of Fig.~\ref{fig:Fe_P04}(a) and the value $T_{1/2}=416 (29)$~ms was obtained. The intense line at $694$~keV originates from the decay of $^{67}$Co ($T_{1/2}=425 (25)$~ms), as reported in Ref.~\cite{Wei_PRC_99}. A laser-enhanced $492$-keV transition is observed in the singles $\gamma$-spectrum of Fig.~\ref{fig:Fe_P04}(b); its absence in the $\beta$-gated spectrum of Fig.~\ref{fig:Fe_P04}(a) indicates the presence of an isomeric transition beyond the $350$-ns time window of the $\beta$-$\gamma$ gate. Its similar half-life behavior compared to the $189$ ($^{67}$Fe decay) and $694$-keV line ($^{67}$Co decay), all about $0.5$~s, did not allow to conclude in which nucleus this transition occurs. However, as the transition is also observed when the lasers are tuned to ionize cobalt, it can not be placed in $^{67}$Fe.

\begin{figure}
\centering
\includegraphics[width=\linewidth]{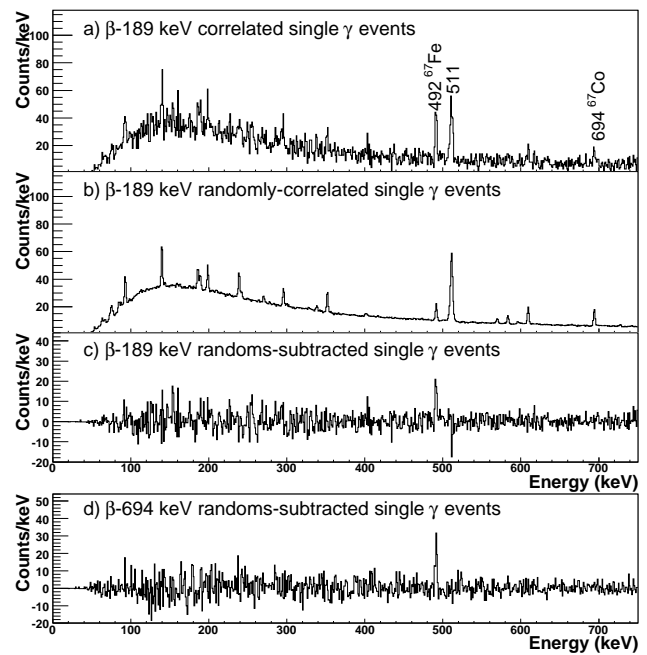}
 \caption{Single $\gamma$-events coming \underline{after} $\beta$-$189$-keV events in a $1 \ \mu$s to $200$ ms window (a), the corresponding randomly-correlated events (b) and randoms-subtracted histogram (c). The randoms-subtracted histogram in (d) corresponds to single $\gamma$-events \underline{before} $\beta$-$694$-keV events in a $1 \ \mu$s to $200$ ms window.}
 \label{fig:Fe_CorrSpectra}
\end{figure}

\begin{figure}
\centering
\includegraphics[width=\linewidth]{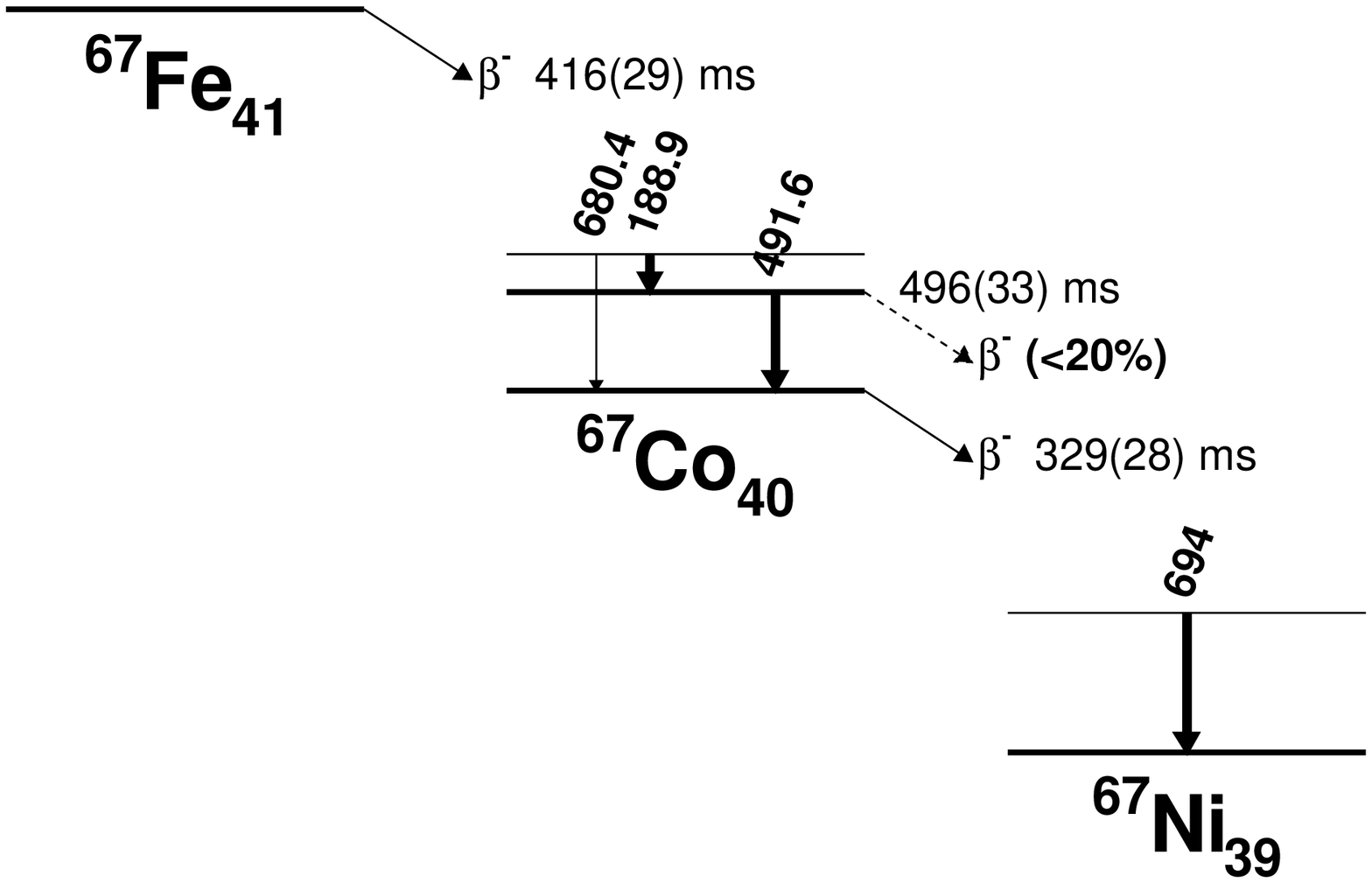}
 \caption{Partial $A=67$ $\beta$-decay chain.}
 \label{fig:Mass67_chain_PRL}
\end{figure}

In order to determine the placement of these newly identified gamma-rays, a new correlation technique \cite{Pau_NIM_08} was developed that takes advantage of the use of free-running digital electronics for the collection of the data. The isomeric $492$-keV transition can be correlated in time with respect to well-established $\gamma$-transitions, like the $189$-keV in $^{67}$Co and the $694$-keV in $^{67}$Ni. Fig.~\ref{fig:Fe_CorrSpectra}(a) shows all correlated single $\gamma$ events in a time window from $1 \ \mu$s to $200$~ms \underline{after} a $\beta$-gated $189$-keV event. After subtraction of the corresponding randomly-correlated events, shown in Fig.~\ref{fig:Fe_CorrSpectra}(b), only the random-free events remain in the spectrum of Fig.~\ref{fig:Fe_CorrSpectra}(c). The latter spectrum contains the $492$-keV line, indicating that the transition takes place after the $\beta$-gated $189$-keV transition. Combined with the presence of the $492$-keV line in Fig. \ref{fig:Fe_CorrSpectra}(d), showing all the random-free events correlated in a time window $1 \ \mu$s to $200$~ms \underline{before} a $\beta$-gated $694$-keV event, the isomeric transition can firmly be placed in $^{67}$Co. The $680$-keV cross-over transition, see Fig.~\ref{fig:Mass67_chain_PRL}, finally establishes an isomeric $492$-keV state.

The half-life of this newly established isomeric state in $^{67}$Co was determined by two independent methods: the $492$-keV $\gamma$-ray decay behavior in a data set with the lasers tuned on cobalt, revealing a half-life of $503(42)$~ms and by fitting the decay behavior of correlated $492$-keV events after $\beta$-gated $189$-keV trigger events in the iron data set, revealing a half-life of $483(56)$~ms. These results indicate the reliability of the half-life values obtained from the correlation technique \cite{Pau_NIM_08}. The half-life value of $496 (33)$~ms, shown in Fig.~\ref{fig:Mass67_chain_PRL}, is the weighted average of the two methods. Additionally, from the correlated $492$-keV events after $\beta$-gated $189$-keV trigger events, the isomeric state at $492$~keV is found to decay by a lower limit of $80 \ \%$ through the $492$-keV $\gamma$-transition, leaving room for at maximum $20 \ \%$ $\beta$-decay. The half-life of the $^{67}$Co ground state was obtained with the same correlation technique as described in \cite{Pau_NIM_08}.

\begin{figure}
\centering
\includegraphics[width=\linewidth]{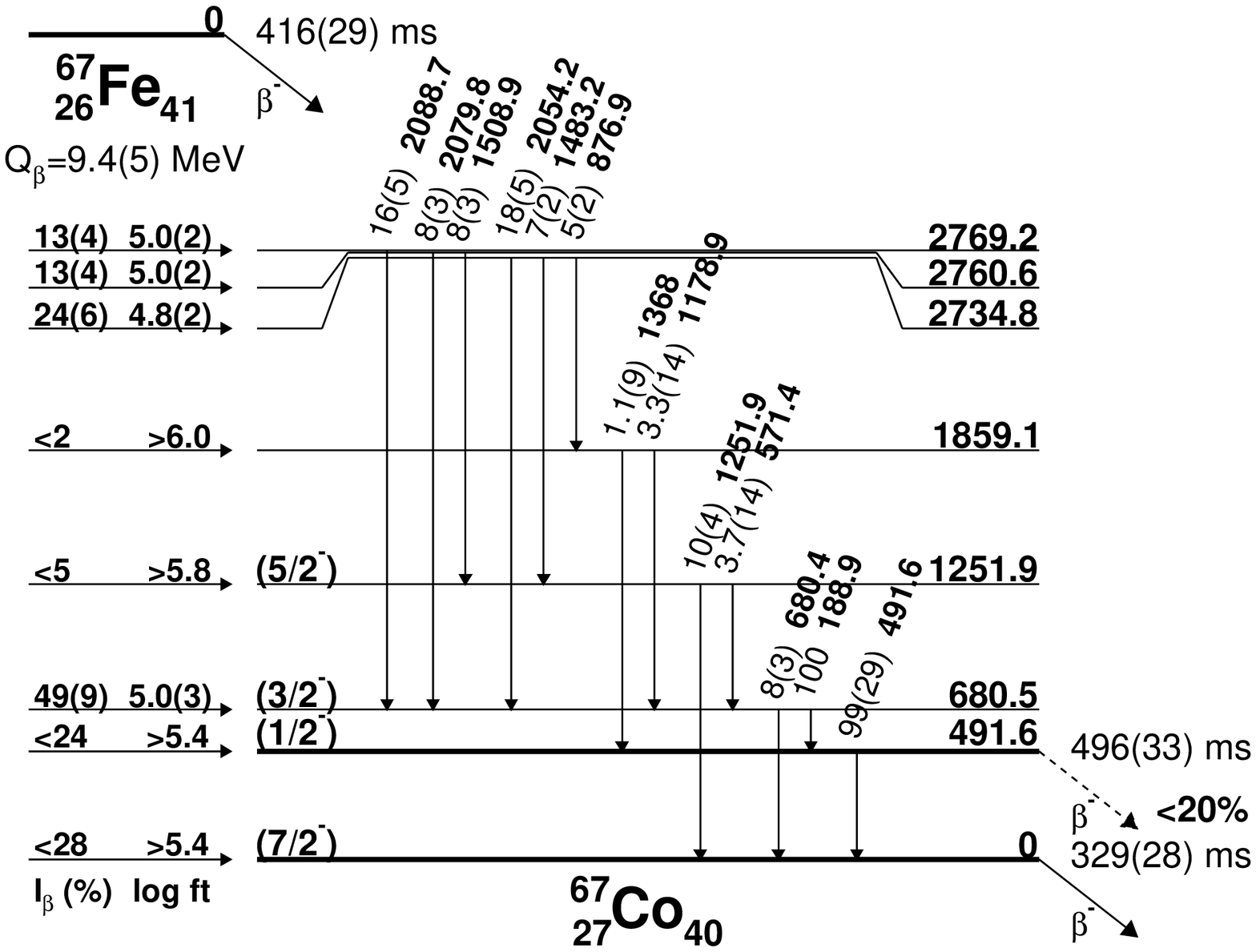}
 \caption{$^{67}$Fe $\beta$-decay scheme. The $\gamma$-ray intensities are indicated relative to the $189$~keV transition ($I_{\gamma}=100$). The upper limits of the $\beta$-decay feeding ($^{67}$Fe to $^{67}$Co) are $1\sigma$-limits on the missed $\gamma$-ray intensities. The log $ft$ values are lower limits due to possibly missed $\gamma$-rays. The upper limit of $\beta$-decay out of the $492$-keV isomer is deduced from the correlations.}
 \label{fig:67Co_scheme}
\end{figure}

The level scheme for $^{67}$Co, shown in Fig.~\ref{fig:67Co_scheme}, has been constructed from $\beta\gamma\gamma$-coincidences. The $\gamma$-ray intensities in $^{67}$Co are relative to the $189$-keV transition ($I_{\gamma}=100$). Assigned spins and parities are based on a $7/2^-$ ground state of $^{67}$Co \cite{Wei_PRC_99}. Weisskopf estimates for the half-life of an E$3$, M$3$ and E$4$ $492$-keV transition are $0.63$~ms, $33$~ms and $500$~s, respectively. Hence, the extracted half-life indicates an M$3$ multipolarity for the $492$-keV transition, which, in turn, leads to $(1/2^-)$ proposed spin and parity values for the $492$-keV isomer in $^{67}$Co. The expected probability ratio for an M$1$ $189$-keV to an E$2$ $680$-keV transition is the only possible match that is of the same order of magnitude as their intensity ratio. Hence, $(3/2^{-})$ spin and parity are proposed for the $680$-keV level and, $(5/2^{-})$ for the $1252$-keV level on the basis of the observed decay to both the $(7/2^{-})$ ground state and $(3/2^{-})$ level at $680$~keV. The $^{67}$Fe decay pattern supports a low spin for its ground state.

\begin{figure}
\centering
\includegraphics[width=\linewidth]{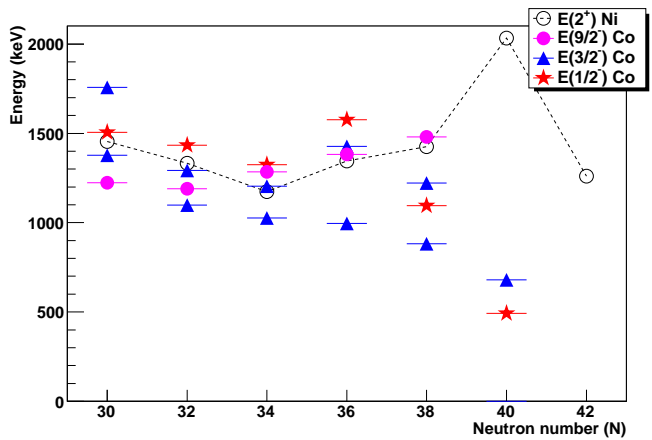}
 \caption{Odd-mass cobalt systematics of low-lying $1/2^{-}$, $3/2^{-}$ and $9/2^{-}$ states relative to the $7/2^{-}$ ground state \cite{NNDC}. The $2^{+}$ energy in the even nickel isotones are indicated by open circles \cite{NNDC}.}
 \label{fig:OddCo_Systematics}
\end{figure}

The level structures for the ground state and low-energy $1/2^{-}$, $3/2^{-}$ and $9/2^{-}$ levels in odd-mass $^{57-67}$Co nuclei are shown in Fig.~\ref{fig:OddCo_Systematics}, along with the energy of the $2^{+}_{1}$ level in the adjacent even-even nickel core nucleus. The data on mass $65$ were obtained from another experiment at LISOL \cite{Pau_PRC_08} and the results are consistent with previous studies \cite{Gau_The_05,Blo_PRL_08}. Higher-spin levels in $^{65}$Co were deduced using the same data set and methods of Ref.~\cite{Hot_PRC_06}.

As can be seen, the energies of the $9/2^{-}$ levels, representative also for the $11/2^{-}$ levels that are not shown in the figure, follow the energies of the core $2^{+}$ levels quite closely. This behavior provides strong support for the description of these nuclei as a single $\pi f_{7/2}^{-1}$ hole in the $Z = 28$ closed shell coupled to excited levels that arise from core excitations~\cite{Reg_PRC_96}. Two $3/2^{-}$ levels and a single $1/2^{-}$ level are also shown that have both core-coupled and $\pi p_{3/2}^{+1}$ and $\pi p_{1/2}^{+1}$ configurations, respectively, as deduced from proton transfer~\cite{Ros_PR_67,Bla_PR_65,Ste_PRC_71} and Coulomb excitation studies~\cite{Nor_NPA_67,Gom_PRC_72}. Note that a $1/2^{-}$ core-coupled state can only be obtained by coupling a $\pi f^{-1}_{7/2}$ hole to a $4^{+}$ state.

All four of these levels follow the energy trend of the core levels up through $N = 34$.  As additional neutrons are added, first the $3/2^{-}$ level drops in energy for $N = 36$ and beyond and the $1/2^{-}$ level for $N = 38$ and particularly $N = 40$ where the core $2^{+}$ and $4^{+}$ energies are $2.033$ and $3.149$~MeV, respectively. Hence, particularly for the $1/2^{-}$ level, core-coupled configuration admixtures should be negligible for $^{67}$Co, leaving proton $\pi (\mathrm{1p-2h})$ excitations across $Z=28$ as the only possible configuration. The strong decrease in excitation energy of the $1/2^{-}$ state can be described by strong proton-neutron correlations inducing deformation \cite{Hey_PR_83}. According to calculated Nilsson orbitals presented in Ref.~\cite{Mol_ADN_97}, a prolate-deformed $^{67}$Co nucleus with a quadrupole deformation value of  $0.25<\varepsilon_2<0.4$ leads in a very natural way to a first excited $[321]1/2^-$ state obtained by promoting one proton particle from the $f_{7/2}$ into the $p_{3/2}$ orbital. Also the neutrons favor this configuration due to the sharply downsloping $1/2^+[440]$ and $3/2^+[431]$ orbitals. It is tempting to assign the $(3/2^{-})$ and $(5/2^{-})$ states at $680$ and $1252$~keV as the first members of a rotational band built on the $(1/2^{-})$ state. Such a rotational band occurs in the indium isotopes~\cite{Hey_PR_83}, where the Coriolis decoupling is even strong enough to bring the $3/2^+$ state below the $\pi [431]1/2^+$ band head. Fig.~\ref{fig:OddCo_Systematics} also indicates that already in $^{65}$Co the $(1/2^{-})$ intruder level sets in at $1095$~keV~\cite{Pau_PRC_08}, $482$~keV lower in energy than the first excited $(1/2^{-})$ level in $^{63}$Co, while the corresponding Ni $4_{1}^{+}$ state goes up by $575$~keV.

In the $N=49$ isotones counterpart, the intruder configuration is at a minimal excitation energy at $_{34}$Se, mid-shell between $Z=28$ and $Z=40$ and goes to a maximum again towards $_{40}$Zr, consistent with a sub-shell closure at $Z=40$ \cite{Hey_PR_83,Mey_PRC_82}. The situation is completely different in the odd-mass cobalt isotopes as the $(1/2^{-})$ proton intruder state is lowest at $N=40$, evidencing that for the $Z=27$ isotopes the $N=40$ sub-shell gap is washed out. In the lighter odd-mass cobalt isotopes the intruder configuration is more difficult to localize due to the strong fragmentation of its strength over different states. It is, remarkably enough, the semi-magic behavior of the $^{68}$Ni core that allows a pure character of the intruder state not mixed up with core-coupled configurations. The rapid onset of deformation below $Z=28$ can thus be explained by the strong proton-neutron residual interactions between the protons in the $\pi f_{7/2}$ orbital and neutrons in the $\nu f_{5/2}$ and $\nu g_{9/2}$ orbitals \cite{Han_PRL_99,Sor_EPJ_03}.

Combining the energy measured for this newly identified proton intruder in $^{67}$Co with that of the previously known $7/2^{-}$ $\pi(\mathrm{2p-1h})$ intruder state in $^{69}$Cu at $1711$~keV~\cite{Oro_NPA_00,Ste_PRL_08}, allows to estimate the energy of the $0^+$ $\pi(\mathrm{2p-2h})$ intruder state in $^{68}$Ni using the prescription of Ref.~\cite{Van_PRL_84}.
At $Z=50$ and $Z=82$ the estimated $0_{2}^{+}$ energies from summing the intruder excitation energies of the $Z \pm 1$ isotones are reproduced on average within $\sim 200$~keV. The $2.2$~MeV estimate for $^{68}$Ni makes the $(0^{+}_{3})$ state at $2.511$~MeV therefore a good candidate for the $\pi(\mathrm{2p-2h})$ configuration. This state was observed in $\beta$-decay work and could not be described by shell model calculations not allowing excitations across $Z=28$ \cite{Mue_PRC_00}, while it was properly predicted by Hartree-Fock-Bogoliubov calculations from collective excitations \cite{Gir_PRC_88}. The low-lying $^{68}$Ni structure is thus dominated by excitations across the $N=40$ ($0^{+}_{2}$ state at $1770$~keV) \cite{Kan_PRC_06} and the $Z=28$ ($0^{+}_{3}$ state at $2511$~keV) (sub)shell gap.

In conclusion, a new correlation technique allowed to identify a $(1/2^{-})$ $496 (33)$ ms isomeric state, decaying by emission of a $492$~keV $\gamma$-ray to the $^{67}$Co ground state. This newly established isomer has been interpreted as a prolate $([321]1/2^{-})$ proton intruder state coexisting with a spherical $(7/2^{-})$ ground state. Taking away only one proton from $^{68}$Ni already induces the obliteration of the $N=40$ sub-shell gap and sets in a region of deformation below $Z=28$. The identification and further study of intruder states in heavier cobalt, nickel and copper isotopes beyond $N=40$ will deliver crucial information on the $Z=28$ gap towards $^{78}$Ni.

We acknowledge the support by the European Commission within the Sixth Framework Programme through I3-EURONS (contract no. RII3-CT-2004-506065), BriX-IUAP P6/23, FWO-Vlaanderen (Belgium), GOA/2004/03 and the Foundation for Polish
Science (AK).


\begin{thebibliography}{33}
\expandafter\ifx\csname natexlab\endcsname\relax\def\natexlab#1{#1}\fi
\expandafter\ifx\csname bibnamefont\endcsname\relax
  \def\bibnamefont#1{#1}\fi
\expandafter\ifx\csname bibfnamefont\endcsname\relax
  \def\bibfnamefont#1{#1}\fi
\expandafter\ifx\csname citenamefont\endcsname\relax
  \def\citenamefont#1{#1}\fi
\expandafter\ifx\csname url\endcsname\relax
  \def\url#1{\texttt{#1}}\fi
\expandafter\ifx\csname urlprefix\endcsname\relax\def\urlprefix{URL }\fi
\providecommand{\bibinfo}[2]{#2}
\providecommand{\eprint}[2][]{\url{#2}}

\bibitem[{\citenamefont{Heyde et~al.}(1983)\citenamefont{Heyde, {P. Van
  Isacker}, Waroquier, Wood, and Meyer}}]{Hey_PR_83}
\bibinfo{author}{\bibfnamefont{K.}~\bibnamefont{Heyde}},
  \bibinfo{author}{\bibnamefont{{P. Van Isacker}}},
  \bibinfo{author}{\bibfnamefont{M.}~\bibnamefont{Waroquier}},
  \bibinfo{author}{\bibfnamefont{J.~L.} \bibnamefont{Wood}}, \bibnamefont{and}
  \bibinfo{author}{\bibfnamefont{R.~A.} \bibnamefont{Meyer}},
  \bibinfo{journal}{Phys. Rep.} \textbf{\bibinfo{volume}{102}},
  \bibinfo{pages}{291} (\bibinfo{year}{1983}).

\bibitem[{\citenamefont{Wood et~al.}(1992)\citenamefont{Wood, Heyde,
  Nazarewicz, Huyse, and {Van Duppen}}}]{Woo_PR_92}
\bibinfo{author}{\bibfnamefont{J.~L.} \bibnamefont{Wood}},
  \bibinfo{author}{\bibfnamefont{K.}~\bibnamefont{Heyde}},
  \bibinfo{author}{\bibfnamefont{W.}~\bibnamefont{Nazarewicz}},
  \bibinfo{author}{\bibfnamefont{M.}~\bibnamefont{Huyse}}, \bibnamefont{and}
  \bibinfo{author}{\bibfnamefont{P.}~\bibnamefont{{Van Duppen}}},
  \bibinfo{journal}{Phys. Rep.} \textbf{\bibinfo{volume}{215}},
  \bibinfo{pages}{101} (\bibinfo{year}{1992}).

\bibitem[{\citenamefont{Broda et~al.}(1995)}]{Bro_PRL_95}
\bibinfo{author}{\bibfnamefont{R.}~\bibnamefont{Broda}} \bibnamefont{et~al.},
  \bibinfo{journal}{Phys. Rev. Lett.} \textbf{\bibinfo{volume}{74}},
  \bibinfo{pages}{868} (\bibinfo{year}{1995}).

\bibitem[{\citenamefont{Stefanescu et~al.}(2008)}]{Ste_PRL_08}
\bibinfo{author}{\bibfnamefont{I.}~\bibnamefont{Stefanescu}}
  \bibnamefont{et~al.}, \bibinfo{journal}{Phys. Rev. Lett.}
  \textbf{\bibinfo{volume}{100}}, \bibinfo{pages}{112502}
  (\bibinfo{year}{2008}).

\bibitem[{\citenamefont{{J. Van Roosbroeck} et~al.}(2004)}]{Van_PRC_04}
\bibinfo{author}{\bibnamefont{{J. Van Roosbroeck}}} \bibnamefont{et~al.},
  \bibinfo{journal}{Phys. Rev. C} \textbf{\bibinfo{volume}{69}},
  \bibinfo{pages}{034313} (\bibinfo{year}{2004}).

\bibitem[{\citenamefont{Mueller et~al.}(1999)}]{Mue_PRL_99}
\bibinfo{author}{\bibfnamefont{W.~F.} \bibnamefont{Mueller}}
  \bibnamefont{et~al.}, \bibinfo{journal}{Phys. Rev. Lett.}
  \textbf{\bibinfo{volume}{83}}, \bibinfo{pages}{3613} (\bibinfo{year}{1999}).

\bibitem[{\citenamefont{Weissman et~al.}(1999)}]{Wei_PRC_99}
\bibinfo{author}{\bibfnamefont{L.}~\bibnamefont{Weissman}}
  \bibnamefont{et~al.}, \bibinfo{journal}{Phys. Rev. C}
  \textbf{\bibinfo{volume}{59}}, \bibinfo{pages}{2004} (\bibinfo{year}{1999}).

\bibitem[{\citenamefont{Sorlin et~al.}(2002)}]{Sor_PRL_02}
\bibinfo{author}{\bibfnamefont{O.}~\bibnamefont{Sorlin}} \bibnamefont{et~al.},
  \bibinfo{journal}{Phys. Rev. Lett.} \textbf{\bibinfo{volume}{88}},
  \bibinfo{pages}{092501} (\bibinfo{year}{2002}).

\bibitem[{\citenamefont{Ameil et~al.}(1998)}]{Ame_EPJ_98}
\bibinfo{author}{\bibfnamefont{F.}~\bibnamefont{Ameil}} \bibnamefont{et~al.},
  \bibinfo{journal}{Eur. Phys. J. A} \textbf{\bibinfo{volume}{1}},
  \bibinfo{pages}{275} (\bibinfo{year}{1998}).

\bibitem[{\citenamefont{Sorlin et~al.}(2000)}]{Sor_NPA_00}
\bibinfo{author}{\bibfnamefont{O.}~\bibnamefont{Sorlin}} \bibnamefont{et~al.},
  \bibinfo{journal}{Nucl. Phys. A} \textbf{\bibinfo{volume}{669}},
  \bibinfo{pages}{351} (\bibinfo{year}{2000}).

\bibitem[{\citenamefont{Facina et~al.}(2004)}]{Fac_NIM_04}
\bibinfo{author}{\bibfnamefont{M.}~\bibnamefont{Facina}} \bibnamefont{et~al.},
  \bibinfo{journal}{Nucl. Instr. and Meth. B} \textbf{\bibinfo{volume}{226}},
  \bibinfo{pages}{401} (\bibinfo{year}{2004}).

\bibitem[{\citenamefont{Eberth et~al.}(2001)}]{Ebe_NPA_01}
\bibinfo{author}{\bibfnamefont{J.}~\bibnamefont{Eberth}} \bibnamefont{et~al.},
  \bibinfo{journal}{Prog. in Part. and Nucl. Phys.}
  \textbf{\bibinfo{volume}{46}}, \bibinfo{pages}{389} (\bibinfo{year}{2001}).

\bibitem[{\citenamefont{Pauwels et~al.}(Accepted)}]{Pau_NIM_08}
\bibinfo{author}{\bibfnamefont{D.}~\bibnamefont{Pauwels}} \bibnamefont{et~al.},
  \bibinfo{journal}{Nucl. Instr. Meth. B}  (\bibinfo{year}{Accepted}).

\bibitem[{\citenamefont{Pauwels et~al.}()}]{Pau_PRC_08}
\bibinfo{author}{\bibfnamefont{D.}~\bibnamefont{Pauwels}} \bibnamefont{et~al.},
  \bibinfo{note}{to be published}.

\bibitem[{\citenamefont{Gaudefroy}(2005)}]{Gau_The_05}
\bibinfo{author}{\bibfnamefont{L.}~\bibnamefont{Gaudefroy}}, Ph.D. thesis,
  \bibinfo{school}{Universit{\'e} de Paris XI Orsay} (\bibinfo{year}{2005}).

\bibitem[{\citenamefont{Block et~al.}(2008)}]{Blo_PRL_08}
\bibinfo{author}{\bibfnamefont{M.}~\bibnamefont{Block}} \bibnamefont{et~al.},
  \bibinfo{journal}{Phys. Rev. Lett.} \textbf{\bibinfo{volume}{100}},
  \bibinfo{pages}{132501} (\bibinfo{year}{2008}).

\bibitem[{\citenamefont{Hoteling et~al.}(2006)}]{Hot_PRC_06}
\bibinfo{author}{\bibfnamefont{N.}~\bibnamefont{Hoteling}}
  \bibnamefont{et~al.}, \bibinfo{journal}{Phys. Rev. C}
  \textbf{\bibinfo{volume}{74}}, \bibinfo{pages}{064313}
  (\bibinfo{year}{2006}).

\bibitem[{\citenamefont{{P. H. Regan} et~al.}(1996)\citenamefont{{P. H. Regan},
  {J. W. Arrison}, {U. J. H{\"u}ttmeier}, and {D. P. Balamuth}}}]{Reg_PRC_96}
\bibinfo{author}{\bibnamefont{{P. H. Regan}}},
  \bibinfo{author}{\bibnamefont{{J. W. Arrison}}},
  \bibinfo{author}{\bibnamefont{{U. J. H{\"u}ttmeier}}}, \bibnamefont{and}
  \bibinfo{author}{\bibnamefont{{D. P. Balamuth}}}, \bibinfo{journal}{Phys.
  Rev. C} \textbf{\bibinfo{volume}{54}}, \bibinfo{pages}{1084}
  (\bibinfo{year}{1996}).

\bibitem[{\citenamefont{Rosner and Holbrow}(1967)}]{Ros_PR_67}
\bibinfo{author}{\bibfnamefont{B.}~\bibnamefont{Rosner}} \bibnamefont{and}
  \bibinfo{author}{\bibfnamefont{C.~H.} \bibnamefont{Holbrow}},
  \bibinfo{journal}{Phys. Rev.} \textbf{\bibinfo{volume}{154}},
  \bibinfo{pages}{1080} (\bibinfo{year}{1967}).

\bibitem[{\citenamefont{Blair and Armstrong}(1965)}]{Bla_PR_65}
\bibinfo{author}{\bibfnamefont{A.~G.} \bibnamefont{Blair}} \bibnamefont{and}
  \bibinfo{author}{\bibfnamefont{D.~D.} \bibnamefont{Armstrong}},
  \bibinfo{journal}{Phys. Rev.} \textbf{\bibinfo{volume}{140}},
  \bibinfo{pages}{B1567} (\bibinfo{year}{1965}).

\bibitem[{\citenamefont{{{K. W. C. Stewart}, {B. Castel} and {B. P.
  Singh}}}(1971)}]{Ste_PRC_71}
\bibinfo{author}{\bibnamefont{{{K. W. C. Stewart}, {B. Castel} and {B. P.
  Singh}}}}, \bibinfo{journal}{Phys. Rev. C} \textbf{\bibinfo{volume}{4}},
  \bibinfo{pages}{2131} (\bibinfo{year}{1971}).

\bibitem[{\citenamefont{Nordhagen et~al.}(1967)\citenamefont{Nordhagen, Elbek,
  and Herskind}}]{Nor_NPA_67}
\bibinfo{author}{\bibfnamefont{R.}~\bibnamefont{Nordhagen}},
  \bibinfo{author}{\bibfnamefont{B.}~\bibnamefont{Elbek}}, \bibnamefont{and}
  \bibinfo{author}{\bibfnamefont{B.}~\bibnamefont{Herskind}},
  \bibinfo{journal}{Nucl. Phys. A} \textbf{\bibinfo{volume}{104}},
  \bibinfo{pages}{353} (\bibinfo{year}{1967}).

\bibitem[{\citenamefont{{Jos{\'e} M. G. G{\'o}mez}}(1972)}]{Gom_PRC_72}
\bibinfo{author}{\bibnamefont{{Jos{\'e} M. G. G{\'o}mez}}},
  \bibinfo{journal}{Phys. Rev. C} \textbf{\bibinfo{volume}{6}},
  \bibinfo{pages}{149} (\bibinfo{year}{1972}).

\bibitem[{\citenamefont{M{\"o}ller et~al.}(1997)}]{Mol_ADN_97}
\bibinfo{author}{\bibfnamefont{P.}~\bibnamefont{M{\"o}ller}}
  \bibnamefont{et~al.}, \bibinfo{journal}{At. Data Nucl. Data Tables}
  \textbf{\bibinfo{volume}{66}}, \bibinfo{pages}{131} (\bibinfo{year}{1997}).

\bibitem[{\citenamefont{Meyer et~al.}(1982)\citenamefont{Meyer, Lien, and
  Henry}}]{Mey_PRC_82}
\bibinfo{author}{\bibfnamefont{R.~A.} \bibnamefont{Meyer}},
  \bibinfo{author}{\bibfnamefont{O.~G.} \bibnamefont{Lien}}, \bibnamefont{and}
  \bibinfo{author}{\bibfnamefont{E.~A.} \bibnamefont{Henry}},
  \bibinfo{journal}{Phys. Rev. C} \textbf{\bibinfo{volume}{25}},
  \bibinfo{pages}{682} (\bibinfo{year}{1982}).

\bibitem[{\citenamefont{Hannawald et~al.}(1999)}]{Han_PRL_99}
\bibinfo{author}{\bibfnamefont{M.}~\bibnamefont{Hannawald}}
  \bibnamefont{et~al.}, \bibinfo{journal}{Phys. Rev. Lett.}
  \textbf{\bibinfo{volume}{82}}, \bibinfo{pages}{1391} (\bibinfo{year}{1999}).

\bibitem[{\citenamefont{Sorlin et~al.}(2003)}]{Sor_EPJ_03}
\bibinfo{author}{\bibfnamefont{O.}~\bibnamefont{Sorlin}} \bibnamefont{et~al.},
  \bibinfo{journal}{Eur. Phys. J. A} \textbf{\bibinfo{volume}{16}},
  \bibinfo{pages}{55} (\bibinfo{year}{2003}).

\bibitem[{NND()}]{NNDC}
\bibinfo{note}{URL: \url{http://www.nndc.bnl.gov/ensdf/}}.

\bibitem[{\citenamefont{Oros-Peusquens and Mantica}(2000)}]{Oro_NPA_00}
\bibinfo{author}{\bibfnamefont{A.~M.} \bibnamefont{Oros-Peusquens}}
  \bibnamefont{and} \bibinfo{author}{\bibfnamefont{P.~F.}
  \bibnamefont{Mantica}}, \bibinfo{journal}{Nucl. Phys. A}
  \textbf{\bibinfo{volume}{669}}, \bibinfo{pages}{81} (\bibinfo{year}{2000}).

\bibitem[{\citenamefont{{Van Duppen} et~al.}(1984)\citenamefont{{Van Duppen},
  Coenen, Deneffe, Huyse, Heyde, and {Van Isacker}}}]{Van_PRL_84}
\bibinfo{author}{\bibfnamefont{P.}~\bibnamefont{{Van Duppen}}},
  \bibinfo{author}{\bibfnamefont{E.}~\bibnamefont{Coenen}},
  \bibinfo{author}{\bibfnamefont{K.}~\bibnamefont{Deneffe}},
  \bibinfo{author}{\bibfnamefont{M.}~\bibnamefont{Huyse}},
  \bibinfo{author}{\bibfnamefont{K.}~\bibnamefont{Heyde}}, \bibnamefont{and}
  \bibinfo{author}{\bibfnamefont{P.}~\bibnamefont{{Van Isacker}}},
  \bibinfo{journal}{Phys. Rev. Lett.} \textbf{\bibinfo{volume}{52}},
  \bibinfo{pages}{1974} (\bibinfo{year}{1984}).

\bibitem[{\citenamefont{Mueller et~al.}(2000)}]{Mue_PRC_00}
\bibinfo{author}{\bibfnamefont{W.~F.} \bibnamefont{Mueller}}
  \bibnamefont{et~al.}, \bibinfo{journal}{Phys. Rev. C}
  \textbf{\bibinfo{volume}{61}}, \bibinfo{pages}{054308}
  (\bibinfo{year}{2000}).

\bibitem[{\citenamefont{Girod et~al.}(1988)\citenamefont{Girod, Dessagne,
  Bernas, Langevin, Pougheon, and Roussel}}]{Gir_PRC_88}
\bibinfo{author}{\bibfnamefont{M.}~\bibnamefont{Girod}},
  \bibinfo{author}{\bibfnamefont{P.}~\bibnamefont{Dessagne}},
  \bibinfo{author}{\bibfnamefont{M.}~\bibnamefont{Bernas}},
  \bibinfo{author}{\bibfnamefont{M.}~\bibnamefont{Langevin}},
  \bibinfo{author}{\bibfnamefont{F.}~\bibnamefont{Pougheon}}, \bibnamefont{and}
  \bibinfo{author}{\bibfnamefont{P.}~\bibnamefont{Roussel}},
  \bibinfo{journal}{Phys. Rev. C} \textbf{\bibinfo{volume}{37}},
  \bibinfo{pages}{2600} (\bibinfo{year}{1988}).

\bibitem[{\citenamefont{Kaneko et~al.}(2006)\citenamefont{Kaneko, Hasegawa,
  Mizusaki, and Sun}}]{Kan_PRC_06}
\bibinfo{author}{\bibfnamefont{K.}~\bibnamefont{Kaneko}},
  \bibinfo{author}{\bibfnamefont{M.}~\bibnamefont{Hasegawa}},
  \bibinfo{author}{\bibfnamefont{T.}~\bibnamefont{Mizusaki}}, \bibnamefont{and}
  \bibinfo{author}{\bibfnamefont{Y.}~\bibnamefont{Sun}},
  \bibinfo{journal}{Phys. Rev. C} \textbf{\bibinfo{volume}{74}},
  \bibinfo{pages}{024321} (\bibinfo{year}{2006}).

\end{thebibliography}
\end{document}